\documentclass[amsmath,aps,showpacs,a4paper,10pt]{revtex4}

 \usepackage{epsf}
 \usepackage{graphicx}    

\begin{document}

 \newcommand{\be}[1]{\begin{equation}\label{#1}}
 \newcommand{\ee}{\end{equation}}
 \newcommand{\bea}{\begin{eqnarray}}
 \newcommand{\eea}{\end{eqnarray}}
 \def\disp{\displaystyle}

 \def\gsim{ \lower .75ex \hbox{$\sim$} \llap{\raise .27ex \hbox{$>$}} }
 \def\lsim{ \lower .75ex \hbox{$\sim$} \llap{\raise .27ex \hbox{$<$}} }

 \begin{titlepage}

 \begin{flushright}
 arXiv:1008.4968
 \end{flushright}

 \title{\Large \bf Cosmological Evolution of Quintessence and
 Phantom with a~New Type of Interaction in Dark Sector}

 \author{Hao~Wei\,}
 \email[\,email address:\ ]{haowei@bit.edu.cn}
 \affiliation{Department of Physics, Beijing Institute
 of Technology, Beijing 100081, China}

 \begin{abstract}\vspace{1cm}
 \centerline{\bf ABSTRACT}\vspace{2mm}
 In the present work, motivated by the work of Cai and Su
 [Phys.\ Rev.\  D {\bf 81}, 103514 (2010)], we propose a new
 type of interaction in dark sector, which can change its sign
 when our universe changes from deceleration to acceleration.
 We consider the cosmological evolution of quintessence and
 phantom with this type of interaction, and find that there
 are some scaling attractors which can help to alleviate the
 cosmological coincidence problem. Our results also show that
 this new type of interaction can bring new features to cosmology.
 \end{abstract}


 \maketitle

 \end{titlepage}

 \renewcommand{\baselinestretch}{1.1}


\section{Introduction}\label{sec1}

In dark energy cosmology (see e.g.~\cite{r1} for reviews), the
 cosmological coincidence problem is one of the well-known
 conundrums, which asks: why are we living in an epoch in
 which the densities of dark energy and matter are comparable?
 To alleviate this coincidence problem, it is natural to
 consider the possible interaction between dark energy and
 dark matter in the literature (see
 e.g.~\cite{r2,r3,r4,r5,r6,r7,r8,r9,r10,r11,r17}). In fact,
 since the nature of both dark energy and dark matter is still
 unknown, there is no physical argument to exclude the possible
 interaction between them. On the contrary, some observational
 evidences of the interaction in dark sector have been found
 recently. For instance, Bertolami {\it et al.}~\cite{r12}
 showed that the Abell Cluster A586 exhibits evidence of the
 interaction between dark energy and dark matter, and they
 argued that this interaction might imply a violation of the
 equivalence principle. On the other hand,
 Abdalla {\it et~al.}~\cite{r13} found the signature of
 interaction between dark energy and dark matter by using
 optical, X-ray and weak lensing data from the relaxed galaxy
 clusters. So, it is reasonable to consider the interaction
 between dark energy and dark matter in cosmology.

In the literature, it is usual to assume that dark energy and
 dark matter interact through a coupling term $Q$, according to
 \bea
 &&\dot{\rho}_m+3H\rho_m=Q\,,\label{eq1}\\
 &&\dot{\rho}_{de}+3H(\rho_{de}+p_{de})=-Q\,,\label{eq2}
 \eea
 where $\rho_m$ and $\rho_{de}$ are the densities of dark
 matter and dark energy (we assume that the baryon component
 can be ignored); $p_{de}$ is the pressure of dark energy;
 $H\equiv\dot{a}/a$ is the Hubble parameter; $a$ is the scale
 factor; a dot denotes the derivative with respect to cosmic
 time $t$. Notice that Eqs.~(\ref{eq1}) and (\ref{eq2})
 preserve the total energy conservation equation
 $\dot{\rho}_{tot}+3H(\rho_{tot}+p_{tot})=0$,
 where $\rho_{tot}=\rho_m+\rho_{de}$. Since there is no
 natural guidance from fundamental physics on the interaction
 $Q$, one can only discuss it to a phenomenological level.
 The familiar interactions extensively considered in the
 literature (see e.g.~\cite{r2,r3,r4,r5,r6,r7,r8,r9,r10,r11})
 include $Q=3\alpha H\rho_m$, $Q=3\beta H\rho_{tot}$,
 and $Q=3\eta H\rho_{de}$.

Recently, Cai and Su~\cite{r14} investigated the interaction
 in a way independent of specific interaction forms by using
 the latest observational data. They divided the whole range
 of redshift $z$ into a few bins and set the interaction term
 $\delta(z)=Q/(3H)$ to be a constant in each redshift bin. From
 the latest observational data, they found that $\delta(z)$ is
 likely to cross the non-interacting line ($\delta=0$), namely,
 the sign of interaction $Q$ changed in the approximate
 redshift range of $0.45\,\lsim\, z\,\lsim\, 0.9$. In fact,
 this result raises a remarkable problem. Indeed, most
 interactions extensively considered in the literature, such as
 $Q=3\alpha H\rho_m$, $Q=3\beta H\rho_{tot}$ and
 $Q=3\eta H\rho_{de}$, are always positive or negative and
 hence cannot give the possibility to change their signs. As
 noted by the authors of~\cite{r14}, some new interaction forms
 should be proposed to address this problem.

In the present work, we are interested to propose such a type
 of interaction and consider its implications to cosmology.
 The authors of~\cite{r14} found that the sign of interaction
 $Q$ changed in the approximate redshift range of
 $0.45\,\lsim\, z\,\lsim\, 0.9$. We note that this redshift
 range is coincident with the one of our universe changing
 from deceleration to acceleration~\cite{r1}. So, a simple idea
 naturally comes to our mind. If the interaction $Q$ is
 proportional to the deceleration parameter
 \be{eq3}
 q\equiv -\frac{\ddot{a}}{aH^2}=-1-\frac{\dot{H}}{H^2}\equiv s-1\,,
 \ee
 the sign of $Q$ can change when our universe changes
 from deceleration ($q>0$) to acceleration ($q<0$). Noting that
 the deceleration parameter $q$ is dimensionless, from
 Eqs.~(\ref{eq1}) and (\ref{eq2}), $Q\propto q\dot{\rho}$ and
 $Q\propto qH\rho$ are both viable from the dimensional point
 of view. To be general, we consider the linear combination of
 these two, namely
 \be{eq4}
 Q=q(\alpha\dot{\rho}+3\beta H\rho)\,,
 \ee
 where $\alpha$ and $\beta$ are both dimensionless constants.
 It is not surprising to find $\dot{\rho}$ in the interaction
 $Q$. We note that in the literature (see e.g.~\cite{r10}) the
 derivatives of energy density have already been allowed
 to appear in the general forms of $Q$. However, the key
 point of our new interaction is the deceleration parameter $q$
 in $Q$, which makes our proposal different from the previous
 works. This new feature gives the possibility that interaction
 $Q$ can change its sign, and hence brings some interesting
 results to cosmology. In this work, we would like to consider
 three interactions of this type, namely
 \bea
 &&Q=q(\alpha\dot{\rho}_m+3\beta H\rho_m)\,,\label{eq5}\\
 &&Q=q(\alpha\dot{\rho}_{tot}+3\beta H\rho_{tot})\,,\label{eq6}\\
 &&Q=q(\alpha\dot{\rho}_{de}+3\beta H\rho_{de})\,.\label{eq7}
 \eea

In the present work, we consider the cosmological evolution of
 quintessence and phantom with the above type of interaction.
 In Sec.~\ref{sec2}, we present the dynamical system of
 interacting quintessence and phantom. In
 Secs.~\ref{sec3}---\ref{sec5}, we discuss the cases with $Q$
 given in Eqs.~(\ref{eq5})---(\ref{eq7}), respectively. In
 Sec.~\ref{sec6}, a~brief conclusion is drawn.


\section{Dynamical system of interacting quintessence and phantom}\label{sec2}

In this work, we consider a flat Friedmann-Robertson-Walker
 (FRW) universe. The Friedmann and Raychaudhuri equations are
 given by
 \bea
 &&H^2=\frac{\kappa^2}{3}\rho_{tot}=
 \frac{\kappa^2}{3}(\rho_{de}+\rho_m)\,,\label{eq8}\\
 &&\dot{H}=-\frac{\kappa^2}{2}(\rho_{tot}+p_{tot})=
 -\frac{\kappa^2}{2}(\rho_m+\rho_{de}+p_{de})\,,\label{eq9}
 \eea
 where $\kappa^2\equiv 8\pi G$. The role of dark energy is
 played by quintessence or phantom, namely
 \bea
 &&\rho_{de}=\rho_\phi=
 \frac{1}{2}\epsilon\dot{\phi}^2+V(\phi)\,,\label{eq10}\\
 &&p_{de}=p_\phi=
 \frac{1}{2}\epsilon\dot{\phi}^2-V(\phi)\,,\label{eq11}
 \eea
 in which $\epsilon=+1$ (quintessence) or $\epsilon=-1$
 (phantom); $V(\phi)$ is the potential. In this work, we
 consider the exponential potential
 \be{eq12}
 V(\phi)=V_0\,e^{-\lambda\kappa\phi}\,,
 \ee
 where $\lambda$ is a dimensionless constant. Without loss of
 generality, we choose $\lambda$ to be positive, since we can
 make it positive through field redefinition $\phi\to -\phi$
 if $\lambda$ is negative.

We consider the cosmological evolution of interacting
 quintessence (phantom) by using the method of dynamical
 system~\cite{r15}. Following~\cite{r2,r3,r4,r6}, we introduce
 the following dimensionless variables
 \be{eq13}
 x\equiv\frac{\kappa\dot{\phi}}{\sqrt{6}H}\,,~~~~~~~
 y\equiv\frac{\kappa\sqrt{V}}{\sqrt{3}H}\,,~~~~~~~
 z\equiv\frac{\kappa\sqrt{\rho_m}}{\sqrt{3}H}\,.
 \ee
 With the help of Eqs.~(\ref{eq8})---(\ref{eq11}), the
 evolution equations~(\ref{eq1}) and (\ref{eq2}) can then be
 rewritten as a dynamical system, namely
 \bea
 &&x^\prime=(s-3)x+\frac{1}{\epsilon}\left(
 \sqrt{\frac{3}{2}}\lambda y^2-Q_1\right),\label{eq14}\\
 &&y^\prime=sy-\sqrt{\frac{3}{2}}\lambda xy\,,\label{eq15}\\
 &&z^\prime=\left(s-\frac{3}{2}\right)z+Q_2\,,\label{eq16}
 \eea
 where
 \be{eq17}
 Q_1\equiv\frac{\kappa\,Q}{\sqrt{6}H^2\dot{\phi}}\,,~~~~~~~
 Q_2\equiv\frac{z\,Q}{2H\rho_m}\,,
 \ee
 a prime denotes derivative with respect to the so-called
 $e$-folding time $N\equiv\ln a$, and
 \be{eq18}
 s\equiv-\frac{\dot{H}}{H^2}=3\epsilon x^2+\frac{3}{2}z^2.
 \ee
 The Friedmann constraint equation~(\ref{eq8}) becomes
 \be{eq19}
 \epsilon x^2+y^2+z^2=1\,.
 \ee
 The fractional energy densities of dark energy and dark matter
 are given by
 \be{eq20}
 \Omega_{de}=\Omega_\phi\equiv\frac{\kappa^2\rho_\phi}{3H^2}
 =\epsilon x^2+y^2\,,~~~~~~~
 \Omega_m\equiv\frac{\kappa^2\rho_m}{3H^2}=z^2\,.
 \ee

Once the interaction $Q$ is given, we can obtain  the critical
 points $(\bar{x},\bar{y},\bar{z})$ of the autonomous system
 Eqs.~(\ref{eq14})---(\ref{eq16}) by imposing the conditions
 $\bar{x}^\prime=\bar{y}^\prime=\bar{z}^\prime=0$. Of course,
 they are subject to the Friedmann constraint Eq.~(\ref{eq19}),
 namely, $\epsilon\bar{x}^2+\bar{y}^2+\bar{z}^2=1$. Note that
 these critical points must satisfy the conditions
 $\bar{y}\geq 0$ and $\bar{z}\geq 0$ by definition~(\ref{eq13}),
 and the requirement of $\bar{x},\,\bar{y},\,\bar{z}$ all being
 real. Then, we can discuss the existence and stability of
 these critical points. An attractor is one of the stable
 critical points of the autonomous system.

To study the stability of the critical points of
 Eqs.~(\ref{eq14})---(\ref{eq16}), we substitute the linear
 perturbations $x\to\bar{x}+\delta x$, $y\to\bar{y}+\delta y$
 and $z\to\bar{z}+\delta z$ about the critical point
 $(\bar{x},\bar{y},\bar{z})$ into
 Eqs.~(\ref{eq14})---(\ref{eq16}) and linearize them. Because
 of the Friedmann constraint Eq.~(\ref{eq19}), there are only
 two independent evolution equations, namely
 \bea
 &&\delta x^\prime=(\bar{s}-3)\delta x+\bar{x}\delta s+
 \frac{1}{\epsilon}\left(\sqrt{6}\lambda\bar{y}\delta y
 -\delta Q_1\right)\,,\label{eq21}\\
 &&\delta y^\prime=\bar{y}\delta s+\bar{s}\delta y
 -\sqrt{\frac{3}{2}}\lambda\left(\bar{x}\delta y
 +\bar{y}\delta x\right)\,,\label{eq22}
 \eea
 where
 \be{eq23}
 \bar{s}=\frac{3}{2}\left(\epsilon\bar{x}^2-
 \bar{y}^2+1\right)\,,~~~~~~~
 \delta s=3\left(\epsilon\bar{x}\delta x
 -\bar{y}\delta y\right)\,,
 \ee
 and $\delta Q_1$ is the linear perturbation coming from $Q_1$.
 The two eigenvalues of the coefficient matrix of the above
 equations determine the stability of the critical point.

In the following sections, we will study the dynamics of
 quintessence (phantom) with the interaction $Q$
 given in Eqs.~(\ref{eq5})---(\ref{eq7}), respectively.


\section{The case of $Q=q(\alpha\dot{\rho}_m+3\beta H\rho_m)$}\label{sec3}

Firstly, we consider the case of
 $Q=q(\alpha\dot{\rho}_m+3\beta H\rho_m)$ given in Eq.~(\ref{eq5}).
 Substituting it into Eq.~(\ref{eq1}), one can find that
 \be{eq24}
 \dot{\rho}_m=\frac{\beta q-1}{1-\alpha q}\cdot 3H\rho_m\,.
 \ee
 Then, substituting into Eq.~(\ref{eq5}), we can finally obtain
 \be{eq25}
 Q=\frac{\beta-\alpha}{1-\alpha q}\cdot 3qH\rho_m\,.
 \ee
 At first glance, this interaction form is very similar to the
 familiar $Q=3\eta H\rho_m$ in which $\eta$ is a constant.
 However, the deceleration parameter $q$ in Eq.~(\ref{eq25})
 makes difference. Note that $q=-1-\dot{H}/H^2$ is a variable
 function of time, which changes its sign when the universe
 changes from deceleration to acceleration.


 \begin{table}[tbp]
 \begin{center}
 \begin{tabular}{c|c}
 \hline\hline\\[-4.1mm]
 ~Label~ & Critical Point $(\bar{x},\,\bar{y},\,\bar{z})$\\[0.5mm]
 \hline\\[-4mm]
 M.1p & $+1/\sqrt{\epsilon}\,$,~~~$0\,$,~~~$0\,$\\[1mm]
 M.1m & $-1/\sqrt{\epsilon}\,$,~~~$0\,$,~~~$0\,$\\[1mm]
 M.2p & $+\sqrt{(r_1-r_2-1)/\epsilon}\,$,~~~$0\,$,
 ~~~$\sqrt{2-r_1+r_2}\,$\\[1mm]
 M.2m & $-\sqrt{(r_1-r_2-1)/\epsilon}\,$,~~~$0\,$,
 ~~~$\sqrt{2-r_1+r_2}\,$\\[1mm]
 M.3p & $+\sqrt{(r_1+r_2-1)/\epsilon}\,$,~~~$0\,$,
 ~~~$\sqrt{2-r_1-r_2}\,$\\[1mm]
 M.3m & $-\sqrt{(r_1+r_2-1)/\epsilon}\,$,~~~$0\,$,
 ~~~$\sqrt{2-r_1-r_2}\,$\\[1mm]
 M.4  & $\lambda/(\sqrt{6}\,\epsilon)\,$,~
 ~~$\sqrt{1-\lambda^2/(6\,\epsilon)}\,$,~~~$0\,$\\[1mm]
 M.5  & ~~$\sqrt{3/2}\,(r_1-r_2)/\lambda\,$,
 ~~~$\sqrt{1-r_1+r_2+3\,\epsilon\,(r_1-r_2)^2/(2\lambda^2)}\,$,
 ~~~$\sqrt{r_1-r_2-3\,\epsilon\,(r_1-r_2)^2/\lambda^2}\,$~\\[1mm]
 M.6  & ~~$\sqrt{3/2}\,(r_1+r_2)/\lambda\,$,
 ~~~$\sqrt{1-r_1-r_2+3\,\epsilon\,(r_1+r_2)^2/(2\lambda^2)}\,$,
 ~~~$\sqrt{r_1+r_2-3\,\epsilon\,(r_1+r_2)^2/\lambda^2}\,$~\\[1.29mm]
 \hline\hline
 \end{tabular}
 \end{center}
 \caption{\label{tab1} Critical points for the case of
 $Q=q(\alpha\dot{\rho}_m+3\beta H\rho_m)\,$.}
 \end{table}


Substituting Eq.~(\ref{eq25}) into Eq.~(\ref{eq17}), we find
 that the corresponding $Q_1$ and $Q_2$ are given by
 \be{eq26}
 Q_1=\frac{3}{2}\cdot\frac{(\beta-\alpha)q}{1-\alpha q}
 \cdot\frac{z^2}{x}\,,~~~~~~~
 Q_2=\frac{3z}{2}\cdot\frac{(\beta-\alpha)q}{1-\alpha q}\,.
 \ee
 Notice that $q=s-1$ and $s$ is given in Eq.~(\ref{eq18}).
 Then, substituting them into the autonomous system
 Eqs.~(\ref{eq14})---(\ref{eq16}), we can find the critical
 points and present them in Table~\ref{tab1}. Note that $r_1$
 and $r_2$ are given by
 \be{eq27}
 r_1\equiv\frac{2+2\alpha+3\beta}{6\alpha}\,,~~~~~~~
 r_2\equiv\frac{\sqrt{4\alpha^2+(2+3\beta)^2
 -4\alpha(4+3\beta)}}{6\alpha}\,.
 \ee
 If $\alpha=0$, only the first two critical points (M.1p) and
 (M.1m) in Table~\ref{tab1} can exist, which are trivial
 solutions in fact. If $\alpha\not=0$, for convenience, we can
 regard $r_1$ and $r_2$ as the model-parameters, in place of
 $\alpha$ and $\beta$. By reversing Eq.~(\ref{eq27}), we can
 express $\alpha$ and $\beta$ as functions of $r_1$ and $r_2$,
 namely
 \be{eq28}
 \alpha=\frac{2}{4-12r_1+9r_1^2-9r_2^2}\,,~~~~~~~
 \beta=-\frac{2(2-6r_1+3r_1^2-3r_2^2)}{4-12r_1+9r_1^2-9r_2^2}\,.
 \ee
 Now, we discuss the existence of the critical points in
 Table~\ref{tab1}. Obviously, Points (M.1p) and (M.1m) can
 exist {\em only} for $\epsilon=+1$ (namely quintessence). They
 are both quintessence-dominated solutions, because the
 corresponding $\Omega_m=\bar{z}^2=0$. For Points (M.2p) and
 (M.2m), if $\epsilon=-1$ (namely phantom), we should have
 $r_1-r_2\leq 1$ by requiring $\bar{x}$ to be real. However, in
 this case $\Omega_m=\bar{z}^2\geq 1$ which is physically
 meaningless. Therefore, Points (M.2p) and (M.2m) can exist
 {\em only} for $\epsilon=+1$ (namely quintessence) and
 $1\leq r_1-r_2\leq 2$. Similarly, Points (M.3p) and (M.3m) can
 exist {\em only} for $\epsilon=+1$ (namely quintessence) and
 $1\leq r_1+r_2\leq 2$. For Point (M.4), if $\epsilon=+1$
 (namely quintessence), it can exist under condition
 $\lambda^2\leq 6$; on the other hand, if $\epsilon=-1$ (namely
 phantom), it can exist for any $\lambda$. In fact, it is a
 dark-energy-dominated solution, because the corresponding
 $\Omega_m=\bar{z}^2=0$. Point (M.5) exists under condition
 $1-r_1+r_2+3\,\epsilon\,(r_1-r_2)^2/(2\lambda^2)\geq 0$ and
 $1\geq r_1-r_2-3\,\epsilon\,(r_1-r_2)^2/\lambda^2\geq 0$.
 Point (M.6) exists under condition
 $1-r_1-r_2+3\,\epsilon\,(r_1+r_2)^2/(2\lambda^2)\geq 0$ and
 $1\geq r_1+r_2-3\,\epsilon\,(r_1+r_2)^2/\lambda^2\geq 0$.
 Obviously, Points (M.2p), (M.2m), (M.3p), (M.3m), (M.5) and
 (M.6) are all scaling solutions, because the corresponding
 $\Omega_m=\bar{z}^2\geq 0$.

To study the stability of these critical points, by linearizing
 $Q_1$, we obtain
 \be{eq29}
 \delta Q_1=\frac{3}{2\bar{x}}\cdot\frac{\beta-\alpha}
 {1-\alpha\bar{q}}\cdot\left\{\left(1-\epsilon\bar{x}^2-
 \bar{y}^2\right)\frac{\delta q}{1-\alpha\bar{q}}-\bar{q}\left[
 2\epsilon\bar{x}\delta x+2\bar{y}\delta y+\left(
 1-\epsilon\bar{x}^2-\bar{y}^2\right)\frac{\delta x}{\bar{x}}
 \right]\right\},
 \ee
 where $\bar{q}=\bar{s}-1$ and $\delta q=\delta s$, while
 $\bar{s}$ and $\delta s$ are given in Eq.~(\ref{eq23}).
 Substituting this $\delta Q_1$ into Eqs.~(\ref{eq21})
 and~(\ref{eq22}), the two eigenvalues of the coefficient
 matrix of Eqs.~(\ref{eq21}) and (\ref{eq22}) determine the
 stability of the critical point. In Table~\ref{tab2}, we
 present the eigenvalues for the first 7 critical points in
 Table~\ref{tab1}. For Point (M.1p), noting that its existence
 requires $\epsilon=+1$ (namely quintessence), it can be stable
 under condition
 $(4-4r_1+r_1^2-r_2^2)/(-4r_1+3r_1^2-3r_2^2)\geq 0$ and
 $\lambda\geq\sqrt{6\epsilon}$. For Point (M.1m), noting that
 its existence requires $\epsilon=+1$ (namely quintessence),
 it is unstable because the second eigenvalue is positive
 (nb. $\lambda$ is positive). For Point (M.2p), noting that
 its existence requires $\epsilon=+1$ (namely quintessence) and
 $1\leq r_1-r_2\leq 2$, it can be stable under condition
 $r_2/(2-5r_1+3r_1^2+r_2-3r_2^2)\geq 0$ and
 $r_1-r_2\leq\lambda\sqrt{2(r_1-r_2-1)/(3\epsilon)}$. For
 Point (M.2m), noting that its existence requires $\epsilon=+1$
 (namely quintessence) and $1\leq r_1-r_2\leq 2$, it is
 unstable because the second eigenvalue is positive (nb.
 $\lambda$ is positive). For Point (M.3p), noting that its
 existence requires $\epsilon=+1$ (namely quintessence) and
 $1\leq r_1+r_2\leq 2$, it can be stable under condition
 $r_2/(-2+5r_1-3r_1^2+r_2+3r_2^2)\geq 0$ and
 $r_1+r_2\leq\lambda\sqrt{2(r_1+r_2-1)/(3\epsilon)}$. For
 Point (M.3m), noting that its existence requires $\epsilon=+1$
 (namely quintessence) and $1\leq r_1+r_2\leq 2$, it is
 unstable because the second eigenvalue is positive (nb.
 $\lambda$ is positive). For Point (M.4), noting that its
 existence requires $1-\lambda^2/(6\epsilon)\geq 0$, it can be
 stable under condition $(-9r_1^2\epsilon^2+9r_2^2\epsilon^2+
 6r_1\epsilon\lambda^2-\lambda^4)/\left[\,3(2-4r_1+3r_1^2-
 r_2^2)\epsilon^2-\epsilon\lambda^2\,\right]\leq 0$. Finally,
 the eigenvalues of Points (M.5) and (M.6) are considerably
 involved, and hence we do not present them here. We find
 that they can exist and are stable in proper
 parameter-space~\cite{r16}.

In summary, for the case with interaction
 $Q=q(\alpha\dot{\rho}_m+3\beta H\rho_m)$, we find that
 there are two dark-energy-dominated attractors (M.1p)
 and (M.4), and four scaling attractors (M.2p), (M.3p),
 (M.5) and (M.6). These scaling attractors can help to
 alleviate the cosmological coincidence problem.
 In e.g.~\cite{r3}, it has been found that there is no
 scaling solution in the interacting phantom model with
 the familiar interaction $Q=3\eta H\rho_m$ in which $\eta$
 is a constant. This fact shows that our new interaction
 $Q=q(\alpha\dot{\rho}_m+3\beta H\rho_m)=
 \frac{\beta-\alpha}{1-\alpha q}\cdot 3qH\rho_m$ (nb.
 Eq.~(\ref{eq25})) can bring new results to cosmology.


 \begin{table}[tbp]
 \begin{center}
 \begin{tabular}{c|c}
 \hline\hline\\[-4.1mm]
 ~Point~ & Eigenvalues \\[0.5mm]
 \hline\\[-4mm]
 M.1p & $-3(4-4r_1+r_1^2-r_2^2)/(-4r_1+3r_1^2-3r_2^2)\,$,~~~
 $3-\lambda\sqrt{3/(2\epsilon)}$\\[1mm]
 M.1m & $-3(4-4r_1+r_1^2-r_2^2)/(-4r_1+3r_1^2-3r_2^2)\,$,~~~
 $3+\lambda\sqrt{3/(2\epsilon)}$\\[1mm]
 M.2p & $6r_2(-2+r_1-r_2)/(2-5r_1+3r_1^2+r_2-3r_2^2)\,$,~~~
 $(3/2)\left[\,r_1-r_2-\lambda\sqrt{2(r_1-r_2-1)/(3\epsilon)}
 \,\right]$\\[1.6mm]
 M.2m & $6r_2(-2+r_1-r_2)/(2-5r_1+3r_1^2+r_2-3r_2^2)\,$,~~~
 $(3/2)\left[\,r_1-r_2+\lambda\sqrt{2(r_1-r_2-1)/(3\epsilon)}
 \,\right]$\\[1.6mm]
 M.3p & ~~$6r_2(-2+r_1+r_2)/(-2+5r_1-3r_1^2+r_2+3r_2^2)\,$,~~~
 $(3/2)\left[\,r_1+r_2-\lambda\sqrt{2(r_1+r_2-1)/(3\epsilon)}
 \,\right]$~\\[1.6mm]
 M.3m & ~~$6r_2(-2+r_1+r_2)/(-2+5r_1-3r_1^2+r_2+3r_2^2)\,$,~~~
 $(3/2)\left[\,r_1+r_2+\lambda\sqrt{2(r_1+r_2-1)/(3\epsilon)}
 \,\right]$~\\[1.6mm]
 M.4  & $-3+\lambda^2/(2\epsilon)\,$,~~~
 $(-9r_1^2\epsilon^2+9r_2^2\epsilon^2+6r_1\epsilon\lambda^2
 -\lambda^4)/\left[\,3(2-4r_1+3r_1^2-r_2^2)\epsilon^2-
 \epsilon\lambda^2\,\right]$\\[1mm]
 \hline\hline
 \end{tabular}
 \end{center}
 \caption{\label{tab2} The corresponding eigenvalues for the
 first 7 critical points in Table~\ref{tab1}.}
 \end{table}



\section{The case of $Q=q(\alpha\dot{\rho}_{tot}+3\beta H\rho_{tot})$}\label{sec4}

Here, we consider the case of
 $Q=q(\alpha\dot{\rho}_{tot}+3\beta H\rho_{tot})$ given in
 Eq.~(\ref{eq6}). From Eq.~(\ref{eq8}), it is easy to find
 $\rho_{tot}=3H^2/\kappa^2$. Substituting into Eq.~(\ref{eq6}),
 we can finally obtain
 \be{eq30}
 Q=\frac{6qH^3}{\kappa^2}\left(\frac{3}{2}\beta-\alpha s\right)\,.
 \ee
 Substituting into Eq.~(\ref{eq17}), we find that
 the corresponding $Q_1$ and $Q_2$ are given by
 \be{eq31}
 Q_1=\frac{q}{x}\left(\frac{3}{2}\beta-\alpha s\right)\,,~~~~~~~
 Q_2=\frac{q}{z}\left(\frac{3}{2}\beta-\alpha s\right)\,.
 \ee
 Notice that $q=s-1$ and $s$ is given in Eq.~(\ref{eq18}).
 Then, substituting them into the autonomous system
 Eqs.~(\ref{eq14})---(\ref{eq16}), we can find that there are
 5 critical points and present the first 4 points in
 Table~\ref{tab3}. All the 4 points in Table~\ref{tab3} are
 scaling solutions because $\Omega_m=\bar{z}^2\geq 0$. The
 last Point (T.3) is considerably involved and hence we do not
 present it here, except to mention that it is also a scaling
 solution. Note that $r_3$ and $r_4$ are given by
 \be{eq32}
 r_3\equiv\frac{2-4\alpha+3\beta}{4+6\alpha}\,,~~~~~~~
 r_4\equiv\frac{\sqrt{4-24\alpha+4\alpha^2+
 20\beta-12\alpha\beta+9\beta^2}}{4+6\alpha}\,.
 \ee
 If $4+6\alpha=0$, all the 4 critical points in
 Table~\ref{tab3} cannot exist. If $4+6\alpha\not=0$, for
 convenience, we can regard $r_3$ and $r_4$ as
 the model-parameters, in place of $\alpha$ and $\beta$. By
 reversing Eq.~(\ref{eq32}), we can express $\alpha$ and
 $\beta$ as functions of $r_3$ and $r_4$, namely
 \be{eq33}
 \alpha=-\frac{2(-1+2r_3+3r_3^2-3r_4^2)}{1+
 6r_3+9r_3^2-9r_4^2}\,,~~~~~~~
 \beta=-\frac{2(-1+2r_3+7r_3^2-7r_4^2)}{1+6r_3+9r_3^2-9r_4^2}\,.
 \ee
 Here, we briefly discuss the existence of the critical
 points. For Points (T.1p) and (T.1m), if $\epsilon=-1$ (namely
 phantom), we should have $r_3+r_4\leq 0$ by requiring
 $\bar{x}$ to be real. However, in this case
 $\Omega_m=\bar{z}^2\geq 1$ which is physically meaningless.
 Therefore, Points (T.1p) and (T.1m) can exist {\em only} for
 $\epsilon=+1$ (namely quintessence) and $1\geq r_3+r_4\geq 0$.
 Similarly, Points (T.2p) and (T.2m) can exist {\em only} for
 $\epsilon=+1$ (namely quintessence) and $1\geq r_3-r_4\geq 0$.
 Point (T.3) can exist in proper parameter-space~\cite{r16}.


 \begin{table}[tbp]
 \begin{center}
 \begin{tabular}{c|c}
 \hline\hline\\[-4.1mm]
 ~Label~ & Critical Point $(\bar{x},\,\bar{y},\,\bar{z})$\\[0.5mm]
 \hline\\[-3.89mm]
 T.1p & ~~$+\sqrt{(r_3+r_4)/\epsilon}\,$,~~~$0\,$,
 ~~~$\sqrt{1-r_3-r_4}\,$~\\[1mm]
 T.1m & ~~$-\sqrt{(r_3+r_4)/\epsilon}\,$,~~~$0\,$,
 ~~~$\sqrt{1-r_3-r_4}\,$~\\[1mm]
 T.2p & ~~$+\sqrt{(r_3-r_4)/\epsilon}\,$,~~~$0\,$,
 ~~~$\sqrt{1-r_3+r_4}\,$~\\[1mm]
 T.2m & ~~$-\sqrt{(r_3-r_4)/\epsilon}\,$,~~~$0\,$,
 ~~~$\sqrt{1-r_3+r_4}\,$~\\[1mm]
 \hline\hline
 \end{tabular}
 \end{center}
 \caption{\label{tab3} The first 4 critical points for the case
 of $Q=q(\alpha\dot{\rho}_{tot}+3\beta H\rho_{tot})\,$.}
 \end{table}


To study the stability of these critical points, by linearizing
 $Q_1$, we obtain
 \be{eq34}
 \delta Q_1=\frac{1}{\bar{x}}\left[\left(\frac{3}{2}\beta-
 \alpha\bar{s}\right)\left(\delta q-\frac{\bar{q}}{\bar{x}}
 \delta x\right)-\alpha\bar{q}\delta s\right]\,,
 \ee
 where $\bar{q}=\bar{s}-1$ and $\delta q=\delta s$, while
 $\bar{s}$ and $\delta s$ are given in Eq.~(\ref{eq23}).
 Substituting this $\delta Q_1$ into Eqs.~(\ref{eq21})
 and~(\ref{eq22}), the two eigenvalues of the coefficient
 matrix of Eqs.~(\ref{eq21}) and (\ref{eq22}) determine the
 stability of the critical point. In Table~\ref{tab4}, we
 present the eigenvalues for the first 4 critical points in
 Table~\ref{tab3}. For Point (T.1p), noting that its existence
 requires $\epsilon=+1$ (namely quintessence) and
 $1\geq r_3+r_4\geq 0$, it can be stable under condition
 $\frac{-6(r_3-r_4)(r_3+r_4)^3-12r_3(r_3
 +r_4)\epsilon^2+18\epsilon^4}{(1+3r_3-3r_4)(r_3+r_4)(1+3r_3+
 3r_4)\epsilon^2}\leq 0$ and
 $\frac{3}{2}+\frac{3\epsilon^2}{2(r_3+r_4)}-
 \lambda\sqrt{\frac{3\epsilon}{2(r_3+r_4)}}\leq 0$. For Point
 (T.1m), noting that its existence requires $\epsilon=+1$
 (namely quintessence) and $1\geq r_3+r_4\geq 0$, it is
 unstable because the second eigenvalue is positive (nb.
 $\lambda$ is positive). For Point (T.2p), noting that its
 existence requires $\epsilon=+1$ (namely quintessence) and
 $1\geq r_3-r_4\geq 0$, it can be stable under condition
 $\frac{-6(r_3+r_4)(r_3-r_4)^3-
 12r_3(r_3-r_4)\epsilon^2+18\epsilon^4}{(1+3r_3-
 3r_4)(r_3-r_4)(1+3r_3+3r_4)\epsilon^2}\leq 0$ and
 $\frac{3}{2}+\frac{3\epsilon^2}{2(r_3-r_4)}-
 \lambda\sqrt{\frac{3\epsilon}{2(r_3-r_4)}}\leq 0$. For Point
 (T.2m), noting that its existence requires $\epsilon=+1$
 (namely quintessence) and $1\geq r_3-r_4\geq 0$, it is
 unstable because the second eigenvalue is positive (nb.
 $\lambda$ is positive). Finally, the eigenvalues of Point
 (T.3) are considerably involved, and hence we do not present
 them here. We find that it can exist and is stable in proper
 parameter-space~\cite{r16}.

So, in the case of
 $Q=q(\alpha\dot{\rho}_{tot}+3\beta H\rho_{tot})$,
 there are 3 scaling attractors (T.1p), (T.2p) and (T.3).
 These scaling attractors can help to alleviate
 the cosmological coincidence problem. Of course, these
 scaling solutions are also different from the ones in the
 interacting quintessence or phantom model with the usual
 interaction $Q=3\eta H\rho_{tot}$ in which $\eta$ is a
 constant. Our new interaction
 $Q=q(\alpha\dot{\rho}_{tot}+3\beta H\rho_{tot})$ brings
 new results.


 \begin{table}[tbp]
 \begin{center}
 \begin{tabular}{c|c}
 \hline\hline\\[-4.1mm]
 ~Point~ & Eigenvalues \\[0.5mm]
 \hline\\[-3mm]
 T.1p & ~~$\frac{-6(r_3-r_4)(r_3+r_4)^3-12r_3(r_3
 +r_4)\epsilon^2+18\epsilon^4}{(1+3r_3-3r_4)(r_3+r_4)(1+3r_3+
 3r_4)\epsilon^2}\,$,~~~
 $\frac{3}{2}+\frac{3\epsilon^2}{2(r_3+r_4)}-
 \lambda\sqrt{\frac{3\epsilon}{2(r_3+r_4)}}$~\\[1.6mm]
 T.1m & ~~$\frac{-6(r_3-r_4)(r_3+r_4)^3-12r_3(r_3+r_4)
 \epsilon^2+18\epsilon^4}{(1+3r_3-3r_4)(r_3+r_4)(1+3r_3
 +3r_4)\epsilon^2}\,$,~~~
 $\frac{3}{2}+\frac{3\epsilon^2}{2(r_3+r_4)}+
 \lambda\sqrt{\frac{3\epsilon}{2(r_3+r_4)}}$~\\[1.6mm]
 T.2p & ~~$\frac{-6(r_3+r_4)(r_3-r_4)^3-
 12r_3(r_3-r_4)\epsilon^2+18\epsilon^4}{(1+3r_3-
 3r_4)(r_3-r_4)(1+3r_3+3r_4)\epsilon^2}\,$,~~~
 $\frac{3}{2}+\frac{3\epsilon^2}{2(r_3-r_4)}-
 \lambda\sqrt{\frac{3\epsilon}{2(r_3-r_4)}}$~\\[1.6mm]
 T.2m & ~~$\frac{-6(r_3+r_4)(r_3-r_4)^3-12r_3(r_3
 -r_4)\epsilon^2+18\epsilon^4}{(1+3r_3-3r_4)
 (r_3-r_4)(1+3r_3+3r_4)\epsilon^2}\,$,~~~
 $\frac{3}{2}+\frac{3\epsilon^2}{2(r_3-r_4)}+
 \lambda\sqrt{\frac{3\epsilon}{2(r_3-r_4)}}$~\\[2.2mm]
 \hline\hline
 \end{tabular}
 \end{center}
 \caption{\label{tab4} The corresponding eigenvalues for the
 first 4 critical points in Table~\ref{tab3}.}
 \end{table}



\section{The case of $Q=q(\alpha\dot{\rho}_{de}+3\beta H\rho_{de})$}\label{sec5}

In this section, we consider the case of
 $Q=q(\alpha\dot{\rho}_{de}+3\beta H\rho_{de})$ given in
 Eq.~(\ref{eq7}). Substituting it into Eq.~(\ref{eq2}), one
 can find that
 \be{eq35}
 \dot{\rho}_{de}=\frac{-3H}{1+\alpha q}\cdot\left(\rho_{de}
 +p_{de}+\beta q\rho_{de}\right)\,.
 \ee
 Then, substituting into Eq.~(\ref{eq7}), we can finally obtain
 \be{eq36}
 Q=\frac{3Hq}{1+\alpha q}\cdot\left[\,\beta\rho_{de}-\alpha
 \left(\rho_{de}+p_{de}\right)\,\right]\,,
 \ee
 which is valid for any dark energy. In the present work, the
 role of dark energy is played by quintessence or phantom.
 Substituting Eqs.~(\ref{eq10}) and (\ref{eq11})
 into Eq.~(\ref{eq36}), we have
 \be{eq37}
 Q=\frac{3Hq}{1+\alpha q}\cdot\left[\left(\frac{\beta}{2}
 -\alpha\right)\epsilon\dot{\phi}^2+\beta V\right]\,.
 \ee
 Substituting into Eq.~(\ref{eq17}), we find that
 the corresponding $Q_1$ and $Q_2$ are given by
 \be{eq38}
 Q_1=\frac{3q}{2(1+\alpha q)}\cdot\left[\left(\beta-2\alpha
 \right)\epsilon x+\frac{\beta y^2}{x}\right]\,,~~~~~~~
 Q_2=\frac{3q}{2(1+\alpha q)z}\cdot\left[\left(\beta-2\alpha
 \right)\epsilon x^2+\beta y^2\right]\,.
 \ee
 Notice that $q=s-1$ and $s$ is given in Eq.~(\ref{eq18}).
 Then, substituting them into the autonomous system
 Eqs.~(\ref{eq14})---(\ref{eq16}), we can find that there are
 8 critical points and present the first 4 points in
 Table~\ref{tab5}. All the 4 points in Table~\ref{tab5} are
 scaling solutions because $\Omega_m=\bar{z}^2\geq 0$. The
 last 4 Points (D.3), (D.4) (D.5) and (D.6) are considerably
 involved and hence we do not present them here, except to
 mention that they are also scaling solutions. Note that $r_5$
 and $r_6$ are given by
 \be{eq39}
 r_5\equiv\frac{2+4\alpha-3\beta}{6\alpha}\,,~~~~~~~
 r_6\equiv\frac{\sqrt{4\alpha^2+
 4\alpha(10-3\beta)+(2-3\beta)^2}}{6\alpha}\,.
 \ee
 If $\alpha=0$, all the 4 critical points in Table~\ref{tab5}
 cannot exist. If $\alpha\not=0$, for convenience, we can
 regard $r_5$ and $r_6$ as the model-parameters, in place of
 $\alpha$ and $\beta$. By reversing Eq.~(\ref{eq39}), we can
 express $\alpha$ and $\beta$ as functions of $r_5$ and $r_6$,
 namely
 \be{eq40}
 \alpha=\frac{8}{9r_6^2-9r_5^2+6r_5-1}\,,~~~~~~~
 \beta=\frac{2(3r_6^2-3r_5^2-6r_5+5)}{9r_6^2-9r_5^2+6r_5-1}\,.
 \ee
 Here, we briefly discuss the existence of the critical
 points. For Points (D.1p) and (D.1m), if $\epsilon=-1$ (namely
 phantom), we should have $r_6\leq r_5$ by requiring
 $\bar{x}$ to be real. However, in this case
 $\Omega_m=\bar{z}^2\geq 1$ which is physically meaningless.
 Therefore, Points (D.1p) and (D.1m) can exist {\em only} for
 $\epsilon=+1$ (namely quintessence) and $r_6\geq r_5$.
 Similarly, Points (D.2p) and (D.2m) can exist {\em only} for
 $\epsilon=+1$ (namely quintessence) and $r_5+r_6\leq 0$.
 Points (D.3), (D.4), (D.5) and (D.6) can exist in proper
 parameter-space~\cite{r16}.

\newpage 


 \begin{table}[tbp]
 \begin{center}
 \begin{tabular}{c|c}
 \hline\hline\\[-4.1mm]
 ~Label~ & Critical Point $(\bar{x},\,\bar{y},\,\bar{z})$\\[0.5mm]
 \hline\\[-3.89mm]
 D.1p & ~~~~$+\sqrt{(r_6-r_5)/\epsilon}\,$,~~~$0\,$,
 ~~~$\sqrt{1+r_5-r_6}\,$~\\[1mm]
 D.1m & ~~~~$-\sqrt{(r_6-r_5)/\epsilon}\,$,~~~$0\,$,
 ~~~$\sqrt{1+r_5-r_6}\,$~\\[1mm]
 D.2p & ~~$+\sqrt{-(r_5+r_6)/\epsilon}\,$,~~~$0\,$,
 ~~~$\sqrt{1+r_5+r_6}\,$~\\[1mm]
 D.2m & ~~$-\sqrt{-(r_5+r_6)/\epsilon}\,$,~~~$0\,$,
 ~~~$\sqrt{1+r_5+r_6}\,$~\\[1mm]
 \hline\hline
 \end{tabular}
 \end{center}
 \caption{\label{tab5} The first 4 critical points for the case
 of $Q=q(\alpha\dot{\rho}_{de}+3\beta H\rho_{de})\,$.}
 \end{table}


To study the stability of these critical points, by linearizing
 $Q_1$, we obtain
 \be{eq41}
 \delta Q_1=\frac{3}{2(1+\alpha\bar{q})}\cdot\left\{\bar{q}
 \left[\left(\beta-2\alpha\right)\epsilon\delta x-
 \frac{\beta\bar{y}^2}{\bar{x}^2}\delta x+
 \frac{2\beta\bar{y}}{\bar{x}}\delta y\right]+\left[\left(
 \beta-2\alpha\right)\epsilon\bar{x}+\frac{\beta\bar{y}^2}{\bar{x}}
 \right]\cdot\frac{\delta q}{1+\alpha\bar{q}}\right\}\,,
 \ee
 where $\bar{q}=\bar{s}-1$ and $\delta q=\delta s$, while
 $\bar{s}$ and $\delta s$ are given in Eq.~(\ref{eq23}).
 Substituting this $\delta Q_1$ into Eqs.~(\ref{eq21})
 and~(\ref{eq22}), the two eigenvalues of the coefficient
 matrix of Eqs.~(\ref{eq21}) and (\ref{eq22}) determine the
 stability of the critical point. In Table~\ref{tab6}, we
 present the eigenvalues for the first 4 critical points in
 Table~\ref{tab5}. For Point (D.1p), noting that its existence
 requires $\epsilon=+1$ (namely quintessence) and
 $r_6\geq r_5$, it can be stable under condition
 $r_6/[(3r_5-3r_6-1)(1+r_5+r_6)]\geq 0$ and
 $1-r_5+r_6-\lambda\sqrt{2(r_6-r_5)/(3\epsilon)}\leq 0$. For
 Point (D.1m), noting that its existence requires $\epsilon=+1$
 (namely quintessence) and $r_6\geq r_5$, it is
 unstable because the second eigenvalue is positive (nb.
 $\lambda$ is positive). For Point (D.2p), noting that its existence
 requires $\epsilon=+1$ (namely quintessence) and
 $r_5+r_6\leq 0$, it can be stable under condition
 $r_6/[(1-3r_5-3r_6)(1+r_5-r_6)]\geq 0$ and
 $1-r_5-r_6-\lambda\sqrt{-2(r_5+r_6)/(3\epsilon)}\leq 0$. For
 Point (D.2m), noting that its existence requires $\epsilon=+1$
 (namely quintessence) and $r_5+r_6\leq 0$, it is
 unstable because the second eigenvalue is positive (nb.
 $\lambda$ is positive). Finally, the eigenvalues of Points
 (D.3), (D.4), (D.5) and (D.6) are considerably involved, and
 hence we do not present them here. We find that they can exist
 and are stable in proper parameter-space~\cite{r16}.

So, in the case of
 $Q=q(\alpha\dot{\rho}_{de}+3\beta H\rho_{de})$, there are
 6 scaling attractors (D.1p), (D.2p), (D.3), (D.4), (D.5)
 and (D.6). These scaling attractors can help to alleviate
 the cosmological coincidence problem. Of course, these
 scaling solutions are also different from the ones in the
 interacting quintessence or phantom model with the usual
 interaction $Q=3\eta H\rho_{de}$ in which $\eta$ is a
 constant. Our new interaction
 $Q=q(\alpha\dot{\rho}_{de}+3\beta H\rho_{de})$ brings
 new results.


\section{Conclusion}\label{sec6}

In the present work, motivated by the recent work of
 Cai and Su~\cite{r14}, we proposed a new type of interaction
 in dark sector, which can change its sign when our universe
 changes from deceleration to acceleration. We considered the
 cosmological evolution of quintessence and phantom with this
 type of interaction. We found that there are some scaling
 attractors which can help to alleviate the cosmological
 coincidence problem. Our results also showed that this new
 type of interaction can bring new features to cosmology.


\section*{ACKNOWLEDGEMENTS}
We are grateful to Prof.~Rong-Gen~Cai and Prof.~Shuang~Nan~Zhang
 for helpful discussions. We~also thank Minzi~Feng, as well as
 Qiping~Su, Xiao-Peng~Ma and M.~A.~Kamran, for kind help and
 discussions. This work was supported in part by NSFC under
 Grant No.~10905005, the Excellent Young Scholars Research
 Fund of Beijing Institute of Technology, and the Fundamental
 Research Fund of Beijing Institute of Technology.


 \begin{table}[tbp]
 \begin{center}
 \begin{tabular}{c|c}
 \hline\hline\\[-4.1mm]
 ~Point~ & Eigenvalues \\[0.5mm]
 \hline\\[-3.3mm]
 D.1p & $\frac{24r_6(r_5-r_6)}{(3r_5-3r_6-1)(1+r_5+r_6)}\,$,~~~
 $(3/2)\left[1-r_5+r_6-\lambda\sqrt{2(r_6-r_5)/(3\epsilon)}
 \,\right]$\\[1.6mm]
 D.1m & $\frac{24r_6(r_5-r_6)}{(3r_5-3r_6-1)(1+r_5+r_6)}\,$,~~~
 $(3/2)\left[1-r_5+r_6+\lambda\sqrt{2(r_6-r_5)/(3\epsilon)}
 \,\right]$\\[1.6mm]
 D.2p & ~~$\frac{24r_6(r_5+r_6)}{(1-3r_5-3r_6)(1+r_5-r_6)}\,$,~~~
 $(3/2)\left[1-r_5-r_6-\lambda\sqrt{-2(r_5+r_6)/(3\epsilon)}
 \,\right]$~\\[1.6mm]
 D.2m & ~~$\frac{24r_6(r_5+r_6)}{(1-3r_5-3r_6)(1+r_5-r_6)}\,$,~~~
 $(3/2)\left[1-r_5-r_6+\lambda\sqrt{-2(r_5+r_6)/(3\epsilon)}
 \,\right]$~\\[2.0mm]
 \hline\hline
 \end{tabular}
 \end{center}
 \caption{\label{tab6} The corresponding eigenvalues for the
 first 4 critical points in Table~\ref{tab5}.}
 \end{table}


\renewcommand{\baselinestretch}{1.1}



\begin{thebibliography}{99}
\bibitem{r16}
In fact, the particular parameter-space for the existence
 and/or stability of this critical point is considerably
 involved and verbose. Since our main aim is to point out
 the fact that it can exist and is stable, the corresponding
 parameter-space is not necessary to be presented here.
 One can work it out comfortably with the help of computer,
 especially with the help of Mathematica.
\end{thebibliography}
\end{document}